\documentclass[a4paper,10pt]{article}

\title{ Multiverse in the Third Quantized Formalism}
\author{Mir Faizal\\
Department of Physics and Astronomy,\\
University of Waterloo, Waterloo,\\
Ontario N2L 3G1, Canada}
\date{}
\begin{document}

\maketitle

\begin{abstract}
In this paper we will analyze the third quantization of
 gravity in path integral formalism. 
We will use the time-dependent version of Wheeler-DeWitt 
equation to analyze 
the multiverse in this formalism. We  will propose a mechanism for 
  baryogenesis to occurs  in the multiverse, without 
violating the  baryon number  conservation. 
\end{abstract}

\section{Introduction}
 The cosmological asymmetry between 
matter and antimatter in our universe is thought 
to be caused by the process of baryogenesis.
  In this process, the baryon number is not conserved \cite{baryon}.
In Grand Unified Theories (GUT), the 
baryon number conservation is violated
by interactions of 
gauge bosons and leptoquarks. This forms the basis of all
models of GUT baryogenesis 
 \cite{b2, b4, b5, b6}. 
In this paper we will explain baryogenesis without violating baryon number 
 conservation in the multiverse but only 
effectively  violating it in particular  universes. 
The existence of the multiverse first appeared 
in the many-worlds interpretation
of quantum theory
\cite{1a}. This idea has now resurfaced in the 
landscape of string theory  \cite{3a, 4a}. 
As this landscape is populated by $10^{500}$ 
vacuum states \cite{1h}, the possibility of all of
 them being real vacuum states 
for different universes remains an open one \cite{1w}.   
As a matter of fact, this model of the multiverse 
has also been used as an explanation 
for inflation in  chaotic inflationary multiverse  \cite{2a}. 
It is a well-known fact that 
second quantization is more suited to
 studying many particle systems as compared to first quantization. Similarly,  
third quantization \cite{2ba,3ba} is naturally suited to studying many 
universes i.e., the multiverse 
 \cite{ 81, 9}.
  It may be noted that the third quantization of the Brans-Dicke theories
  \cite{i} and Kaluza-Klein theories  \cite{ia} 
 have been already  thoroughly studied.

In this paper we will first review the generalization of the wave function of the 
universe \cite{hh} to the time-dependent case  \cite{5,6}. 
We will then analyse the multiverse in this model. 
In most model of the multiverse, universes do not interact 
with each other. However, in the models  of multiverse 
that are used to study spacetime foam, baby universes 
branch off from a parent universe \cite{wh1, wh2, wh3, wh4, wh5}.
 These models have been used to study the topological
changes in spacetime, and thus  have been employed  to
analyse the existence
 of the cosmological constant \cite{col, col21, col2, col1}.
Thus, these models are suited to study the formation of our universe from a collision of 
two previous universes.
 In these models, interactions are introduced, and different universes interact with each other
 at Planck scale \cite{st, st1}. 
 In this paper, we will use this model of interacting universes to explain the  occurrence of 
  baryogenesis   without 
violating the  baryon number  conservation. 
This will be done by assuming that an initial universe split into two different universes. 
The sum total of the baryon numbers in both the universes is conserved, but one universe can have 
more matter if the other one has exactly the same amount of extra anti-matter. 
Thus, the baryon number gets violated in individual universes without getting violated in the full 
multiverse. 
 It may be noted a that a similar analysis 
has been performed in  minisuperspace models for Horava-Lifshits gravity 
\cite{mir}. As the third quantized universes can have their own conversed third quantized charges, 
a similar analysis  has  been performed in the fourth quantized formalism \cite{mir1}.

\section{Time Dependent Wheeler-DeWitt Equation}
In this section we will review the time-dependent wave function 
for the universe, which is a solution to the time-dependent Wheeler-DeWitt equation \cite{5,6}.
The wave function of the universe can be obtained from 
the no-boundary proposal as follows \cite{hh} 
\begin{equation}
\Psi[h_{ij},\tilde\phi, \tilde A ]=-\int Dg D\phi  DA \mu[g,\phi, A ]\hbox{exp}(iS[g,\phi, A ]),
\end{equation}
Here, $g$ denotes the metric, $\phi$ denotes all the
 matter fields and $A$ denotes the all
gauge fields. 
 This integral is taken   over all 
geometries with a compact boundary on which the induced metric is 
$h_{ij}$. Furthermore,    the  induced value of the matter and gauge 
fields on this boundary are denoted by 
$\tilde\phi $ and  $ \tilde A$, respectively.  
 The invariance of the 
 measure $\mu$ under an infinitesimal translation of $N$ leads to the invariance of the 
wave function under the same. From this the Wheeler-DeWitt equation follows as
\begin{eqnarray}
{\delta \Psi\over \delta N } &=& -i\int Dg D\phi DA \mu[g,\phi, A]{\delta S
\over \delta N}{\exp}(iS).
\end{eqnarray}
Thus we get, 
\begin{equation}
 H \phi  =0,
\end{equation}
where
\begin{eqnarray}
 H &=& h^{1/2}\left[- h^{-1/2}G_{ijkl}\frac{\delta^2}{ \delta h_{ij}
\delta h_{kl}}+R^{(3)}(h)\right.\nonumber \\ &&\left.
 -2\Lambda -\frac{16\pi}{ M_P^2}
T_{nn}\left(-i\frac{\delta}{ \delta \tilde \phi},
\tilde \phi, \tilde A, -i\frac{\delta}{ \delta \tilde A},
\right)\right]\Psi[h_{ij},\tilde 
\phi, \tilde A]=0.
\end{eqnarray}
Here, $G_{ijkl}$ is the metric on the superspace,
\begin{equation}
G_{ijkl}={1\over 2}h^{-1/2}(h_{ik}h_{jl}+h_{il}h_{jk}-h_{ij}h_{kl}),
\end{equation}
and 
\begin{equation}
K_{ij}=\frac{1}{ N}\big[-\frac{1}{ 2}\frac{\partial h_{ij}}{ \partial t}
+N_{(i| j)}\big].
\end{equation}
The scalar curvature $R^{(3)}$ is constructed from $h_{ij}$, 
and the covariant derivative with respect to the later 
 quantity is represented by a stroke. 
The projection of the 
 stress-energy tensor for  the matter  and  gauge fields normal to the surface
 is denoted by $T_{nn}$. 
We obtain the momentum
constraint by following a similar procedure,   
\begin{eqnarray}
{\delta \Psi\over \delta N_i } &=& -i\int Dg D\phi DA \mu[g,\phi, A]
{\delta S
\over \delta N_i}{\exp}(iS).
\end{eqnarray}
Thus we get, 
\begin{equation}
 H_i \Psi  =0,
\end{equation}

In the multiverse, the wave function of the universe would change. Thus,
  we would expect that the wave
function of the universe to  be analogous to a wave function for the 
 time-dependent 
Schroedinger equation. Therefore, we want to obtain a time-dependent version of the 
Wheeler-DeWitt equation. 
It may be noted that in quantum mechanics a wave function can be constructed by using the 
path integral formalism 
\begin{equation}
\label{pathintegral}
\psi ({\vec x},t)=-\int D{\vec x}(t) \hbox{exp}[iS({\vec x}(t))].
\end{equation}
The Schroedinger equation  can be obtained from this expression for the wave function 
\begin{equation}
i{\partial \psi\over \partial t}=H\psi. 
\end{equation}
This is because, we can write 
\begin{equation}
{\partial \psi\over \partial t}=-i\int D{\vec x}(t)
{\partial S\over \partial t}\hbox{exp}(iS).
\end{equation}

Now, a time-dependent 
version of the Wheeler-DeWitt equation can be constructed by following a similar  procedure.
Thus, the time-dependent wave function of 
the universe can define  as \cite{5,6},
\begin{equation}
\Psi[h_{ij},\tilde\phi, \tilde A ]
=-\int Dg D\phi  DA M[g,\phi, A ]\hbox{exp}(iS[g,\phi, A ]),
\end{equation}
where, again $g$ denotes  the metric, $\phi$ denotes all the matter fields,
  $A$ denotes the all
gauge fields. The 
  integral again is taken over all geometries with 
  a compact boundary. However, now
 the  measure factor  $M[g,\phi, A]$  breaks the time
translational invariance of the path integral.
This in turn makes the  wave function of the universe  
  time-dependent. A possible choice for the measure, which does not change 
  the quantum mechanical equation arising from 
  the momentum constraint,  
  $H_i \Psi =0$, can be written as  
\begin{equation}
M[g,\phi, A ]=\mu [g,\phi, A]N^b.
\end{equation}
This measure bring an explicit time dependence, but retains the 
the invariance of the spatial three-geometry even at quantum level. 
We can now write   
\begin{eqnarray}
{\delta \Psi\over \delta N } &=& -\int Dg D\phi DA 
{\delta  M\over \delta N }
\hbox{exp}(iS)\nonumber \\ && -i\int Dg D\phi DA M[g,\phi, A]{\delta S
\over \delta N}{\exp}(iS).
\end{eqnarray}
The time-dependent version of the Wheeler-DeWitt equation obtained from this 
definition of the wave function can be written as 
\begin{equation}
-i \frac{\delta  \Psi}{ \delta N }+ {H}\Psi =0.
\end{equation}

\section{Third Quantization}

Now the wave function depends on the full metric $\Psi(g, \phi, A)$. So we can write the 
partition function for the multiverse as 
\begin{equation}
 Z_0 = \int D \Psi[g, \phi, A] \exp \left(i  S[\Psi(g,\phi, A)]\right),
\end{equation}
where 
\begin{eqnarray}
 S[\Psi(g, \phi, A )] &=& \int Dg D\phi DA  M[g,\phi, A] \Psi(g,\phi, A) 
\nonumber \\ && \times \left(-i\frac{\delta \Psi}{ \delta N } 
 + {H}\Psi\right)\Psi(g,\phi, A),
\end{eqnarray}
 is a free action for the multiverse. 
We can now define generating function for universes as follows 
\begin{equation}
 Z_0[J] = \int D\Psi \exp( i S[\Psi] + i J \Psi), 
\end{equation}
where 
\begin{equation}
 J\Psi = \int Dg  D\phi DA M[g,\phi, A] J(g,\phi, A) \Psi (g,\phi, A).
\end{equation}
Now if we rotate $N$ in complex plane as $N \to i N$, then we have 
\begin{equation}
 Z_0[J] = \int D\Psi \exp( - S[\Psi] - J \Psi), 
\end{equation}
This can now be expressed as 
\begin{eqnarray}
 Z_0[J] &=&  \int Dg_1 Dg_2  D \phi_1 D \phi_2 DA_1 DA_2  M[g_1, \phi_1,A_1]
M[g_2,\phi_2, A_2]
 \nonumber \\ 
  &&\times\exp (J(g_1,\phi_1, A_1)
 \Delta[(g_1,\phi_1, A_1), (g_2,\phi_2, A_2)]  \nonumber\\
&& \times J (g_2,\phi_2, A_2) ), 
\end{eqnarray}
where 
\begin{equation}
 \Delta[(g_1,\phi_1, A_1), (g_2,\phi_2, A_2)]
 = \left[Det\left(-i \frac{\delta 
 \Psi}{ \delta N }+ {H}\right)\right]^{-1}. 
\end{equation}
Now  we define 
\begin{eqnarray}
&& \left(-i\frac{\delta 
 \Psi}{ \delta N }+ {H} \right)\Delta[(g_1,\phi_1, A_1),
 (g_2,\phi_2, A_2)]  
\nonumber \\ &=& \delta(g_1, g_2)
\delta(\phi_1, \phi_2)
\delta( A_1,  A_2),
\end{eqnarray}
 where 
\begin{eqnarray}
 && \int Dg_1 D\phi_1 DA_1 M[ g_1, \phi_1, A_1]\delta(g_1, g_2) \delta(
\phi_1, \phi_2) \nonumber \\ && \times
\delta(A_1, A_2) \Psi (g_1,\phi_1, A_1) 
\nonumber \\&=& \Psi (g_2, \phi_2, A_2). 
\end{eqnarray}
Now we can written 
\begin{equation}
 \Delta [(g_1,\phi_1, A_1), (g_2,\phi_2, A_2)] = 
\frac{\delta^2 Z[J]}{\delta J(g_1, \phi_1, A_1)
 \delta J(g_2, \phi_2, A_2)}\left.\right|_{J =0}. 
\end{equation}
This is the amplitude for the wave function of a universe to change  in the multiverse. 

In most model of spacetime foam, baby universes interact with the parent universe at the 
Plank scale \cite{col, col21}.
The interaction term for a model of
  interacting universes  has been analysed  
for time-independent Wheeler-DeWitt equation in third quantization 
formalism \cite{st, st1}. We will generalize this model to time-dependent Wheeler-DeWitt equation and 
analyse some natural consequences of it. Thus, following on from the previous work on spacetime 
foam, we add the following
 interaction 
to the free action, 
\begin{eqnarray}
 S_{int} &=& \lambda \int Dg_3Dg_1 Dg_2 D\phi_3 D\phi_1 D\phi_2
 DA_3 DA_1 DA_2\nonumber \\ && M[g_1, \phi_1, A_1]M[g_2, \phi_2, A_2]M[g_3, \phi_3, A_3]
\nonumber \\  && \times
( \Psi(g_1, \phi_1, A_1)  
\Psi(g_2,\phi_2, A_2)\Psi(g_3, \phi_3, A_3)\nonumber \\ &&
 \delta(g-g_1)
\delta(g-g_2)\delta(\phi-\phi_1)
\delta(\phi-\phi_2)\delta(A-A_1)\nonumber \\&&
\delta(A-A_2)  ).
\end{eqnarray}
We can now define generating function for universes as follows 
\begin{equation}
 Z[J] = \int D\psi \exp (- S_t[\Psi] -  J \Psi), 
\end{equation}
where 
\begin{equation}
 S_t [\Psi] = S[\Psi] + S_{int}[\Psi].
\end{equation}
So we have 
\begin{equation}
 Z[J] = \int D\Psi \exp (- S_t[\Psi] -  J \Psi). 
\end{equation}
Now we get
\begin{equation}
 Z[J] = \frac{\exp - S_{int}  Z_0[J] }{(\exp - S_{int}  Z_0[J])\left.\right|
_{J=0} },
\end{equation}
where 
\begin{equation}
 S_{int}[\lambda \Psi^3 ] =
 S_{int}\left[\lambda \left(-\frac{\delta }{\delta J}\right)^{3} \right]. 
\end{equation}
In fact, as we can only observe the universes that interact with our universes, we have to use the
generating function for connected  universes only. This can be written as 
\begin{equation}
 W [J] = - In Z[J]
\end{equation}

We can calculate the amplitude  for  a universe $U_1$
 to 
split into two universes  $U_2$ and $U_3$.
Now if $\Psi(g_1, \phi_1, A_1)$ is the wave function for $U_1$, 
 $\Psi(g_2, \phi_2, A_2)$ is the wave function for $U_2$ and 
 $\Psi(g_3, \phi_3, A_3)$ for $U_3$,
 then this amplitude is given  by  
\begin{eqnarray}
&& G[(g_1,\phi_1 A_1), (g_2, \phi_2, A_2),( g_3, \phi_3, A_3 )] \nonumber \\&=& 
\frac{\delta^3 W[J]}{\delta J(g_1, \phi_1, A_1) \delta J(g_2, \phi_2, A_2)
 \delta J(g_3, \phi_3, A_3)}\left.\right|_{J =0}  \nonumber \\
&=& 
 \lambda \int D g D \phi D A M[g, \phi, A]  
\Delta [(g_1,\phi_1, A_1), (g,\phi, A)] \nonumber \\ && \times 
\Delta [(g,\phi, A), (g_2,\phi_2, A_2)] 
\Delta [(g,\phi, A), (g_3,\phi_3, A_3)] \nonumber \\&& 
+ \mathcal{O} (\lambda^2).
\end{eqnarray}

The 
splitting of $U_1$ into $U_2$ and $U_3$ will appear as a big bang in $U_2$ and $U_3$. 
This can also use used to possibly explain 
the domination of 
matter over antimatter in our universe without violating baryon number conservation \cite{4z}.
Now, if the    baryon number is conserved in  $U_1$,
 then we have 
\begin{equation}
 Bn_1 - Bm_1 =0, 
\end{equation}
where $Bn_1$ represents the total baryon number of the baryons, $Bm_1$ represent the total baryon number of the anti-baryons 
in the universe $U_1$. Here 
$n_1$ are the total number of baryons and $m_1$ are the total number of  anti-baryons in universe 
$U_1$.  Now we represent the total number of baryons and 
 in the universes
 $U_2$ as $n_2,m_2$  and $U_3$ as and $n_3, m_3$, respectively.  The  baryon number 
conservation implies 
\begin{equation}
 Bn_2 + Bn_3 - Bm_2 -Bm_3 =0. 
\end{equation}
However, the baryon number in the multiverse does not constraint   the 
baryon numbers in the universes $U_2$ or $U_3$  to be separately conserved. So, we can write    
\begin{eqnarray}
 Bn_2 - Bm_2 &\neq& 0,\nonumber \\ 
 Bn_3 - Bm_3 &\neq& 0.
\end{eqnarray}
Thus, after splitting of the universe $U_1$,  the universe  $U_2$ can have more matter than 
anti-matter,  if the universe $U_3$ has more 
anti-matter than matter. The extra matter in the $U_2$ will exactly balance the extra anti-matter in 
$U_3$. Thus, the  baryon number will still be conservation in the multiverse. 
\section{Conclusion}
In this paper, we first derive an time-dependent version of the Wheeler-DeWitt equation.
Then we analysed a classical Lagrangian density whose variation generates 
this equation. 
We also analysed the third quantization of this model. 
The occurrence of baryogenesis was explained due 
to the splitting of a earlier universe into two universes. 
 This caused  a effective violation of the baryon number 
conservation 
 in individual universes, but it was still conserved in the  
multiverse. 

It may be noted that we we arbitrary fixed the form of interactions in our model. It would be nice to 
derive the form of interaction from some underlying physical principle. One such model that can be studied 
is a third quantized gauge theory. We can construct the third quantized action from complex third quantized fields
instead of real ones. If we do that, the potential part of the third quantized theory will be invariant under 
a  gauge transformation. However, the kinetic part of the third quantized action can also be made gauge invariant by 
replacing all the second quantized functional derivatives by second quantized covariant functional derivatives. These covariant 
functional derivatives can 
be constructed by defining a third quantized gauge field whose gauge transformations exactly cancel the extra pieces generated 
by the action of the second quantized functional derivatives on the third quantized field, in the third quantized action. 
We can then add a potential term for the third quantized gauge field thus introduced. The third quantized covariant derivatives will 
couple the third quantized fields to these gauge fields, and the interaction thus generated can also account for the baryogenesis. 
It would be interesting to analyse this model further.

\end{document}